\documentclass[titlepage,12pt]{utarticle}
\usepackage{amssymb,amscd}
\usepackage{color}
\usepackage[]{hyperref} 

\numberwithin{equation}{section}

\newcommand{\homoquot}[3]{\raisebox{-.25ex}{\ensuremath{#3}}\backslash\raisebox{.25ex}{\ensuremath{#1}}/\raisebox{-.25ex}{\ensuremath{#2}}}

\newcommand{\ket}[1]{\lvert #1\rangle}

\newcommand{\Tr}{{\rm Tr}}

\newcommand{\ie}{\textit{i.e.},\ }
\newcommand{\etal}{\textit{et al\/}}

\begin{document}
\preprint{
  UTTG--07--03\\
  {\tt hep-th/0312115}\\
}
\title{
  1+1 Dimensional Critical String Theory and Holography
}
\author{Aaron Bergman, Jacques Distler and Uday Varadarajan
    \thanks{Work supported in part by NSF Grant PHY0071512
      and the Robert A.~Welch Foundation.
    }
}
\oneaddress{
      Theory Group, Physics Department\\
      University of Texas at Austin\\
      Austin, TX 78712 USA\\ {~}\\
      \email{abergman@physics.utexas.edu}
      \email{distler@golem.ph.utexas.edu}
      \email{udayv@physics.utexas.edu}
}

\date{December 10, 2003}

\Abstract{
In \href{http://arxiv.org/abs/hep-th/0310120}{hep-th/0310120},
Goheer, Kleban and Susskind argued that the holographic principle
is inconsistent with the existence of stable, Lorentz invariant,
1+1 dimensional compactifications. We note some difficulties with
their analysis and present two novel backgrounds of string theory
with 1+1 noncompact dimensions that satisfy their conditions yet
possess 16 or 24 supersymmetries. It is difficult to believe
that such backgrounds could be unstable.
}

\maketitle


\section{Introduction}\label{sec:intro}
The holographic principle
\cite{tHooft:1993gx,Susskind:1994vu}\footnote{For a review, see
\cite{Bousso:2002ju}.}, loosely the idea that the entropy of an
isolated system should be bounded by its area, suitably defined,
may give us an insight into the structure of the fundamental
theory of quantum gravity. Its genesis was in the observation
that the Hawking temperature of a black hole naturally leads to
an entropy proportional to its horizon area. In string theory,
this conjecture has been verified \cite{Strominger:1996sh} by an
explicit counting of microstates. While the original bound from
the black hole horizon has since been generalized to the
covariant entropy bound \cite{Bousso:1999xy}, it is still of
interest to consider other horizons with associated temperatures.
One immediate example is the Rindler metric which represents
Minkowski space as seen by an accelerated observer. It is known
from Unruh  that this observer sees a temperature of $a/2\pi$.
One might wonder if there is an entropy bound from the horizon in
this metric. This would seem to be in contradiction with the fact
that the Rindler metric is just ordinary Minkowski space in different coordinates. Thankfully, the area of the Rindler horizon is infinite, so one does not obtain any bounds in this situation.

As noticed by Goheer, Kleban and Susskind \cite{Goheer:2003tx}, however, there is one situation where the above argument fails. If we consider a spacetime with only 1+1 noncompact dimensions, then the area of the Rindler horizon is the area of the compact dimensions. In this situation, they argue that the holographic principle provides an entropy bound on the resulting theory. They then argue that the Lorentz algebra generating the symmetries of the noncompact directions is not representable on a Hilbert space in a manner consistent with the finite entropy predicted by the holographic principle. They conclude that 1+1 dimensional compactifications of string theory, which is believed to be holographic, must necessarily be inconsistent in some manner.

In \S\ref{sec:review} and \S\ref{sec:loophole} of this paper, we review the argument
of \cite{Goheer:2003tx} and point out some loopholes in their
reasoning. In \S\ref{sec:rigid}, we
present two Lorentz invariant 1+1 dimensional backgrounds of the
heterotic string that contain no massless moduli in
contradiction with a statement in section 4 of  \cite{Goheer:2003tx}. We also believe that these compactifications are interesting in
their own right. Readers uninterested in holography are invited
to skip to \S\ref{sec:rigid} for the presentation of the model.

\section{A Review of Goheer, Kleban and Susskind}\label{sec:review}
As noted above, the fundamental observation of \cite{Goheer:2003tx} is that the area of the Rindler horizon in a compactification to 1+1 dimensions is finite. The argument proceeds from this fact as follows. As the observer in Rindler coordinates perceives a thermal bath, we begin with a thermal density matrix $\rho = e^{-\beta H}/Z$ where $Z = \Tr\ e^{-\beta H}$. From the holographic principle we then set $S = -\Tr\ \rho\log\rho$ where $S$ is proportional to the volume of the compactified dimensions. We can manipulate the formula as follows (following \cite{Goheer:2003tx}):
\begin{equation} 
\label{sform}
S = (1/Z)\Tr\ e^{-\beta H} (\log Z + \beta H) = \log Z + (\beta / Z) \Tr\ He^{-\beta H}\ .
\end{equation}
As $\rho$ is unaffected by a constant shift in $H$, we can assume without loss of generality that the spectrum of $H$ is positive. We will assume that $H$ is self-adjoint. Then, by the functional calculus (\cite{ReedSimon}, p.~222), the second term in \eqref{sform} is manifestly postive, so we have:
\begin{equation}
S \ge \log \Tr\ e^{-\beta H}\ .
\end{equation}
Thus, $e^{-\beta H}$ must be trace class and, hence, compact. It is also self-adjoint as $H$ is self-adjoint. Therefore, by the Riesz-Schauder theorem (\cite{ReedSimon}, p.~203), the spectrum is a discrete set with only possibly zero as a limit point. Then it follows, again from the functional calculus, that $H$ has a discrete spectrum with no limit points as any such limit point would map to a non-zero limit point in the spectrum of $e^{-\beta H}$.

The remainder of the proof follows from the identification of the
Hamiltonian with the generator of Lorentz boosts. It is shown
that this implies a contradiction with the above assumptions by
way of the the existence of an orthonormal basis of eigenvectors
of $H$ (the Hilbert-Schmidt theorem (\cite{ReedSimon}, p.~203)). We must conclude that one of our assumptions is incorrect. Goheer \etal\ choose the existence of such a compactification as the incorrect assumption.

\section{Compactifications With Moduli and a Possible
Loophole}\label{sec:loophole}
One can, however, examine the other assumptions to escape their
conclusion. We will focus on the application of the holographic
principle. First, let us examine the effective field theory in
1+1 dimensions obtained by Kaluza-Klein reduction on the
compactified dimensions. Let us assume that the volume of the
compactified dimensions is a modulus and hence a scalar field,
$\sigma$. If there were a higher number of noncompact dimensions,
the theory would separate into superselection sectors each
labelled by a value of $\sigma$. This value could then rightfully
be called the volume of the compact dimensions. However, in two
dimensions this is not possible. Infrared fluctuations in two
dimensions cause the wavefunction of the theory to spread out
across the moduli space. In this situation, there is no sharply
defined notion of the volume of the extra dimensions. At best,
this is a quantum operator whose expectation value may or may not be finite. 

We should not, however, conclude \cite{Goheer:2003tx} that, because of the large fluctuations of the scalar field, the theory effectively decompactifies.  In theories of extended supergravity, the scalar fields form a non-linear sigma model with a target space given by $\homoquot{G}{K}{G(\BZ)}$ where $G$ is some split form of a Lie group and $K$ is its maximal compact subgroup. A toy example of such a space is $\homoquot{SL(2,\BR)}{SO(2)}{SL(2,\BZ)}$, which arises as a factor in the compactification on a $T^2$. This space is the usual keyhole domain, $\mathcal{F}$, in the upper half plane, with the Poincar\'e metric. The important facts about moduli spaces of this form are that they are of finite volume and that the probability the scalar field lies in any particular region in the moduli space is proportional to the volume of that region\footnote{The ground state wave function is a square-integrable harmonic function on the moduli space, and hence, by the existence of the Baily-Borel compactification, a constant.}. 

This leads to an interesting phenomenon. On the upper half plane, the Poincar\'e metric is given by
\begin{equation}
  ds^2 = \frac{dx^2 + dy^2}{y^2} \ .
\end{equation}
The expectation value of the field, $\sigma$, is given by
\begin{equation}
\langle\sigma\rangle = \frac{3}{\pi}\int_{\mathcal{F}} \frac{dx\ dy}{y^2} y\ .
\end{equation}
This integral is, of course, log divergent. Nevertheless, the probability of finding the theory in a configuration of volume greater than $a$ will be (for $a>1$)
\begin{equation}
\mathcal{P}(y\ge a)=\frac{3}{\pi} \int_a^\infty \frac{dy}{y^2} =  \frac{3}{\pi a}
\end{equation}
which goes to zero as $a$ gets larger. Thus, while the expectation value of the volume of the compact space diverges, the \emph{probability} of being at large volume is negligibly small, and the theory does not appear to be decompactified in any real sense.

If, in this 1+1 dimensional context, it is legitimate to replace ``the volume of the compact space'' with ``the \emph{expectation value} of the volume of the compact space,'' in the formulation of the holographic bound, then the aforementioned divergence would provide a resolution of the paradox in \cite{Goheer:2003tx}.

It is easy to come up with compactifications of this sort down to 1+1 dimensions with 8, 16 or 32 unbroken supersymmetries. It is hard to imagine that there could be anything else sick in a theory with that much unbroken supersymmetry.

A more challenging case, addressed by \cite{Goheer:2003tx}, is a compactification in which there are no massless moduli. If the moduli are all frozen, then there is presumably a well-defined notion of the volume of the compact space. One might expect that demanding the absence of massless moduli would require a theory with few or no unbroken supersymmetries. It would not be unreasonable to claim that such a theory would either develop a nonzero cosmological constant or would otherwise be unstable. It is argued in \cite{Goheer:2003tx} that their result on holography constitutes a proof that no such vacuum with zero cosmological constant exists.

In \S\ref{sec:rigid}, we will exhibit two Lorentz invariant
string backgrounds with 1+1 noncompact dimensions that possesses
16 or 24 unbroken supersymmetries and no massless scalars. As
with the previous example, we believe it highly unlikely that any
background with so much supersymmetry could cause trouble. These
backgrounds are obtained as asymmetric orbifolds
\cite{Narain:1986qm} of the heterotic string. One might argue that, because of their non-geometric nature, one cannot apply the holographic principle to asymmetric orbifolds or that one does not know what the holographic bound should be. However, asymmetric orbifolds are the most mildly non-geometrical of the multitude of non-geometrical compactifications of string theory. If one cannot assign a meaningful notion of volume, and apply notions of holography, to an asymmetric orbifold, one would despair of being able to do so for a Landau-Ginsburg model or other exotica. Certainly, given the close relation of the asymmetric orbifold to the ordinary orbifold construction from the worldsheet point of view, whatever the correct definition of the ``volume" is, it is surely finite. For this reason, we believe that our examples are apposite.

\section{Moduli-Free Backgrounds in 1+1 Dimensions}\label{sec:rigid}

A toroidal compactification of the heterotic string down to 2
dimensions is specified by an even, self-dual Lorentzian lattice,
$\Lambda$, of signature $(24,8)$\footnote{For a review, see
\cite{Lerche:1988np}.}. Such a model would have 16 unbroken supercharges and many massless moduli. To obtain a rigid theory, we will choose a particular form for $\Lambda$ and construct an asymmetric orbifold of the resulting toroidal compactification.

Let us take $\Lambda = \Lambda_{24}\oplus \Lambda_{E_{8}}$, where $\Lambda_{24}$ is a Niemeier lattice, \ie\ an even self-dual \emph{Euclidean} lattice of dimension 24, and where $\Lambda_{E_{8}}$ is the $E_{8}$ root lattice.
Any choice of Niemeier lattice will serve our purpose as they all, including a number of orbifolds thereof \cite{Schellekens:1993db}, lead to a left-moving holomorphic CFT  with $c=24$. All that is necessary is that we obtain such a CFT with a modular-invariant partition function, $\chi(q)$.

\subsection{Heterotic Monster Module}\label{sec:HeteroticMonster}
For our first model, we choose $\Lambda_{24}$ to be the Leech
lattice and consider a $\BZ_{2}$ quotient, $X^{i}_{L}\to
-X^{i}_{L}$, yielding the famous Monster Module\footnote{For a physics review, see \cite{Dixon:1988qd}.}. The partition function for the left-movers is
\begin{equation}
   \chi_{\mathrm{monster}}(q)= J(q) = \frac{1}{q}+ 196884 q + 21493760 q^{2} + \dots \ .
\end{equation}
Note that there are no massless states from the left-moving sector. The monster module is the unique $c=24$ holomorphic CFT with this property.

The partition function for the right-moving NS sector is (for ease of notation, we temporarily drop the bars, writing everything as a function of $q$, rather than $\overline{q}$)
\begin{equation}
\begin{split}
   \chi_{NS}&= \frac{1}{2} \frac{\Theta_{E_{8}}(\theta_{3}^{4}-\theta_{4}^{4})}{\eta^{12}}\\
   &= 8 + 2048q + 49152q^2 + 614400q^3 +\dots
\end{split}
\end{equation}
where
\begin{equation}
\Theta_{E_8}= \sum_{p\in\Lambda_{E_8}} q^{p^2/2}
\end{equation}
is the theta function of the $E_8$ lattice. The partition function for the right-moving Ramond sector is
\begin{equation}
\begin{split}
   \chi_{R}&= - \frac{1}{2} \frac{\Theta_{E_{8}}\theta_{2}^{4}}{\eta^{12}}\\
   &= - \chi_{NS}\ .
\end{split}
\end{equation}
The full partition function,
\begin{equation}
\begin{split}
   Z_{NS}(q,\overline{q})= \chi_{\mathrm{monster}}(q)\cdot \overline{\chi}_{NS}(\overline{q}) \\
    Z_{R}(q,\overline{q})= \chi_{\mathrm{monster}}(q)\cdot \overline{\chi}_{R}(\overline{q})
\end{split}
\end{equation}
is manifestly spacetime supersymmetric, with 16 unbroken supersymmetries\footnote{The 1+1 dimensional spacetime theory has $(8,8)$ supersymmetry. Since we are working in light-cone gauge, we see only half of the supercharges, those of one particular chirality, as R-sector ground states.}. There are no massless states, as any potential massless state would be projected out by level-matching. Instead, the lowest mass states are $2048\times 196884=403218432$ bosons and an equal number of fermions at the first mass level.

\subsection{Orbifolding the Right-Movers}\label{sec:OrbifoldRight}
We now return to a general $c=24$ holomorphic CFT for the left-movers. To project out the massless states, we instead orbifold the right-movers.
Consider a $\BZ_{2}$ quotient which does not act on the left-movers and acts as $X^{i}_{R}\to -X^{i}_{R}$ and $\psi^{i}\to -\psi^{i}$ on the right-movers. This action is compatible with right-moving worldsheet supersymmetry.

Let us construct the partition function for the right-movers. For ease of notation, we will, again, temporarily drop the bars, writing everything as a function of $q$, rather than $\overline{q}$.

First consider the untwisted sector. The bosons are integer-moded, but we need to combine them into states which are overall even or odd under $\BZ_{2}$. To obtain a $\BZ_{2}$-even state, we can take an even number of oscillators and act on an even combination of lattice momenta, or we can take an odd number of oscillators and act on an odd combination of lattice momenta. For any nonzero $p\in \Lambda_{E_8}$, we can form even and odd combinations $\ket{p}\pm\ket{-p}$; for $p=0$, there is only one state. This gives:
\begin{equation}
\begin{split}
&\frac{(\Theta_{E_{8}}+1)}{2}\frac{1}{2}\left(\frac{1}{\eta^{8}} + \frac{1}{q^{1/3}\prod_{n=1}^{\infty}(1+q^{n})}\right)
+\frac{(\Theta_{E_{8}}-1)}{2}\frac{1}{2}\left(\frac{1}{\eta^{8}} - \frac{1}{q^{1/3}\prod_{n=1}^{\infty}(1+q^{n})}\right)\\
&=\tfrac{1}{4} (\Theta_{E_{8}}+1) \frac{1+\theta_3^{4}\theta_{4}^{4}}{\eta^{8}}+
\tfrac{1}{4} (\Theta_{E_{8}}-1) \frac{1-\theta_3^{4}\theta_{4}^{4}}{\eta^{8}}\\
&=\frac{1}{2\eta^{8}}(\Theta_{E_{8}}+\theta_3^{4}\theta_{4}^{4})\ .
\end{split}
\end{equation}
Similarly, the bosonic states odd under $\BZ_{2}$ give
\begin{equation}
 \frac{1}{2\eta^{8}}(\Theta_{E_{8}}-\theta_3^{4}\theta_{4}^{4})\ .
\end{equation}

In the untwisted NS sector, we project onto odd fermion number  where we work in the convention where the ground state, the NS tachyon, has even fermion number. Since the fermions are odd, we need to combine this with a bosonic state which is odd under the $\BZ_{2}$. Thus, the GSO-projected partition function in the untwisted NS sector is
\begin{subequations}
\begin{equation}\label{chiuns}
 \chi_{uNS}=\left[\frac{1}{2\eta^{8}}(\Theta_{E_{8}}-\theta_3^{4}\theta_{4}^{4})\right]
\left[\frac{1}{2} \frac{\theta_{3}^{4}-\theta_{4}^{4}}{\eta^{4}}\right]\ .
\end{equation}

In the untwisted R sector, we have a choice of sign for the GSO projection. Since the fermions are odd under $\BZ_{2}$, this choice of sign gets correlated with whether the state of the bosons is even or odd under $\BZ_{2}$. This gives the GSO-projected partition function in the untwisted R sector (remembering the overall minus sign for spacetime fermions) as
\begin{equation}
\chi_{uR}=-\left[\frac{1}{2\eta^{8}}(\Theta_{E_{8}}\mp\theta_3^{4}\theta_{4}^{4})\right]
\left[\frac{1}{2} \frac{\theta_{2}^{4}}{\eta^{4}}\right]\ .
\end{equation}

In the twisted sector, the bosons are half-integer moded, and the fermions integer-moded (NS) or half-integer-moded (R). In the NS sector, we GSO project onto states of odd fermion number, which requires that we also project onto states with an odd number of bosonic oscillators so as to obtain something which is overall even under $\BZ_2$. Thus, remembering that there are 16 ``fixed points" in this asymmetric orbifold (the square-root of the number of fixed points on the symmetric $T^8/\BZ_2$ orbifold), the twisted NS partition function is
\begin{equation}\label{chitns}
 \chi_{tNS}=16 \times \left[\frac{1}{32}\frac{\theta_{2}^{4} \theta_{3}^{4}-\theta_{2}^{4}\theta_{4}^{4}}{\eta^8}\right]
\left[\frac{1}{2} \frac{\theta_{2}^{4}}{\eta^{4}}\right]\ .
\end{equation}
In the twisted R sector, we again have a choice of sign for the GSO projection which gets correlated with the sign of the projection on the bosons giving
\begin{equation}
 \chi_{tR}=-16 \times \left[\frac{1}{32}\frac{\theta_{2}^{4} \theta_{3}^{4}\mp\theta_{2}^{4}\theta_{4}^{4}}{\eta^8}\right]
\left[\frac{1}{2} \frac{\theta_{3}^{4}\mp \theta_{4}^{4}}{\eta^{4}}\right]\ .
\end{equation}
\end{subequations}

Notice that, for one choice of GSO projection in the R sector, there is an exact degeneracy of spacetime bosons and fermions at every mass level. For the other sign, there is a discrepancy, but, using the identity $\theta_2\theta_3\theta_4=2\eta^3$, we see that
\begin{equation}
\begin{split}
   -\chi_{uR}(q) &= 8 + \chi_{uNS}(q)\\
   -\chi_{tR}(q) &= 16 + \chi_{tNS}(q)\ .
\end{split}
\end{equation}
In other words, there are 24 extra massless fermions, all of the same chirality.

The existence of extra massless fermions, with chirality opposite to that of the supercharges is completely compatible with \emph{chiral} supersymmetry in spacetime. Changing the sign of the GSO projection in the R sector flips the chirality of the supercharges and hence also that of these massless fermions. The apparent discrepancy in the spectrum between the two choices of GSO projection is an artifact of light-cone gauge. When one identifies $\tau$ with $X^+$, one can only ``see" massless left-moving spacetime fermions but not those that are right-moving. Thus, the choice of GSO projection merely affects our ability to discern the extra fermions, not their actual existence.

The orbifolding has killed half of the supersymmetries, leaving us with chiral $(8,0)$ supersymmetry. Note however that, in light of the identity,
\begin{equation*}
\Theta_{E_8}= \tfrac{1}{2}(\theta_2^8+\theta_3^8+\theta_4^8)\ ,
\end{equation*}
there is also an exact degeneracy between the spectra of spacetime bosons in the untwisted and twisted sectors and, moreover, they are all massive:
\begin{equation}
 \chi_{uNS}(q)=\chi_{tNS}(q)=1024 q + 24567 q^2 + 307200 q^3 + \dots \ .
\end{equation}

If the left-moving $c=24$ CFT has $n_1$ states of conformal weight 1 and $n_2=196884$ states of conformal weight 2, then there are $24n_1$ massless fermions,  $2048 n_2$ massive scalars and $2048 n_2$ massive fermions at the first excited level, and so on.

The degeneracy between the twisted and untwisted sectors is, of course, no accident. In addition to the 8 spacetime supercharges from the untwisted sector that survive the orbifolding, there are an additional 16 supercharges from the twisted sector. Our full theory has chiral $\mathcal{N}=(24,0)$ spacetime supersymmetry. Since some of the supercharges come from the twisted sector, this naturally leads to equal multiplicities of twisted and untwisted sector states.

The spectrum of massive states can be understood quite simply in the light-cone Green-Schwarz formulation of this theory. Let us replace our right-moving ``internal" SCFT with 24 anti-commuting fields, $\Theta^A$ where $A=1,\dots,24$. The contribution to the zero-point energy in this model is such that the ground state is at mass level 1. Quantizing the zero modes, $\Theta^A_0$, yields 4096 states at this mass level, of which half are spacetime bosons and half spacetime fermions. At higher levels, the spectrum of, say, the bosons is given by
\begin{equation}
\chi_{GS(b)}(q)= 2048 q \prod_{n=1}^\infty (1+q^n)^{24}= \frac{1}{2} \frac{\theta_2^{12}}{\eta^{12}} = \chi_{uNS}(q) +\chi_{tNS}(q)
\end{equation}
in exact agreement with equations \eqref{chiuns} and \eqref{chitns}.

\subsection{No Worries}
We now have two constructions of Lorentz-invariant compactifications of string theory to 1+1 dimensions with \emph{no} massless scalars. In \S\ref{sec:HeteroticMonster}, we found a unique compactification with $(8,8)$ supersymmetry. In \S\ref{sec:OrbifoldRight}, we found a large class of compactifications (one for every $c=24$ holomorphic CFT), with $(24,0)$ supersymmetry.

We will reserve a discussion of the physics of these theories for a future work, noting here only that, with 16 or 24 unbroken supersymmetries, they appear to be perfectly sensible. Again, it is hard to imagine what might destabilize them.

Other rigid models can presumably be constructed along the lines suggested in \cite{Harvey:1988da}. The authors of this latter paper briefly note in their conclusions the existence of the model in \S\ref{sec:HeteroticMonster}.

\section{Discussion}
Goheer \etal\ found what appeared to be a sharp contradiction
between holography and the existence of stable, Lorentz-invariant
1+1 dimensional compactifications of string theory. Not wishing
to abandon holography, they argued that some instability must
beset such compactifications. In particular, they must somehow be
driven to decompactify, as holography dictates that the entropy
of Rindler space be bounded by the volume of the compactified
dimensions.  We have argued that, with enough unbroken supersymmetry, it is \emph{highly unlikely} that any such instability could arise.

In theories with massless moduli, however, we have seen that a plausible mechanism exists for reconciling such 1+1 dimensional compactifications with the exigencies of holography. The moduli and, in particular, the ``volume'' of the compact space do not have definite values in the vacuum. At best, we might hope to have a holographic bound based on the expectation value of the volume of the compact space. One can easily have a situation in which the expectation value of the volume diverges, even though the theory is in no sense decompactified as the probability of the volume being large can be arbitrarily small.

However, we (or, more properly, holography) are not out of the woods. In \S\ref{sec:rigid} we pointed out that there exist compactifications down to 1+1 dimensions with lots of supersymmetries and \emph{no} massless moduli. It is not immediately clear what number one should assign as the ``volume'' of the compact space for these asymmetric orbifolds, but whatever it is, it is surely finite, and the paradox of Goheer \etal\ comes back in force.

One possible way out is to acknowledge that perhaps holography is only a semiclassical notion. After all, when one discusses the values of things like areas, be they of spacelike surfaces or of lightsheets, one is necessarily working in a semiclassical regime where these all have a definite value. The semiclassical approximation is clearly not valid in these 1+1 dimensional situations, so perhaps we should not attempt to apply notions of holography to these compactifications.

Either way, string theory is replete with nongeometrical
backgrounds, and if there is one lesson that we have learned from
our study of string theory, it is that ``geometry'' is an
emergent concept, valid only  in some semiclassical regime. If
holography is to be a fundamental principle in string theory, it
will have to generalize such semiclassical
notions as ``area'' to the quantum regime. 

Still, even when semiclassical consideration are
valid, there is something vaguely sinister about the paradox
posed by Goheer \etal. We could very well be living in spatially
compact universe (or, more relevant to the matter at hand, a
universe with two compact spatial dimensions). But, given the
arguments of \cite{Goheer:2003tx}, as reviewed in
\S\ref{sec:review} above, such a universe, \emph{no matter how
large the scale of compactness}, is incompatible with holography.

None of this rings a death knell for holography in two or any
other number of dimensions. Nonetheless, we believe that these examples present
evidence against the strongest conclusions of Goheer \etal. It
may be that the ultimate fate of holography in two dimensions
must await a better understanding of either holography itself or its
implementation in string theory.

\section*{Acknowledgements}
We would like to thank Matthew Kleban and Donald Marolf for useful discussions.

\bibliography{leech}
\bibliographystyle{utphys}

\end{document}